\documentclass[]{article}  %>>> use for US letter paper
\usepackage[]{graphicx}
\newcommand{\be}{\begin{equation}}
\newcommand{\ee}{\end{equation}}
\newcommand{\lo}{{\cal L}_0}
\title{A new generalized differential image motion monitor}

\author{Eric Aristidi, Yan Fante\"{\i}-Caujolle, Aziz Ziad, C\'ecile Dimur, \\ Julien Chab\'e and Baptiste Roland}
\date{
Laboratoire Lagrange UMR 7293 UNS-CNRS-OCA, Parc Valrose, 06108 Nice Cedex 2, France}

\begin{document} 
  %\bigskip
  
  \maketitle 

%%%%%%%%%%%%%%%%%%%%%%%%%%%%%%%%%%%%%%%%%%%%%%%%%%%%%%%%%%%%% 
\section*{abstract}
We present first results of a new instrument, the Generalized Differential Image Motion Monitor (GDIMM), aiming at monitoring parameters of the optical turbulence (seeing, isoplanatic angle, coherence time and outer scale). GDIMM is based on a small telescope equipped with a 3-holes mask at its entrance pupil. The seeing is measured by the classical DIMM technique using two sub-pupils of the mask (6~cm diameter separated by a distance of 20~cm), the isoplanatic angle is estimated from scintillation through the third sub-pupil (its diameter is 10~cm, with a central obstruction of 4~cm). The coherence time is deduced from the temporal structure function of the angle of arrival (AA) fluctuations, thanks to the high-speed sampling rate of the camera. And the difference of the motion variances from sub-apertures of different diameters makes it possible to estimate the outer scale. GDIMM is a compact and portable instrument, and can be remotely controlled by an operator. We show in this paper the first results of test campaigns obtained in 2013 and 2014 at Nice observatory and the Plateau de Calern (France). Comparison with simultaneous data obtained with the Generalized Seeing Monitor (GSM) are also presented. 

%>>>> Include a list of keywords after the abstract 

%%%%%%%%%%%%%%%%%%%%%%%%%%%%%%%%%%%%%%%%%%%%%%%%%%%%%%%%%%%%%
\section{INTRODUCTION}
\label{par:intro}  % \label{} allows reference to this section
Since several years, our group develops original techniques and instrumentation for measuring the optical turbulence of the atmosphere. Several prototypes were developped in the past, such as the generalized seeing monitor (GSM)\cite{Ziad00} which has become a reference for monitoring the coherence parameters of the wavefront at ground level, i.e. the seeing $\epsilon$, isoplanatic angle $\theta_0$, coherence time $\tau_0$ and outer scale $\lo$. In the last 15 years GSM was used in a large number of astronomical observatories and for prospecting potential new sites (see [\cite{Ziad00}] and references therein). 

Since the beginning of the years 2000, our group was also engaged in the site qualification of the site of Dome C in Antarctica. Specific prototypes of a DIMM, an isoplanometer (to monitor  $\theta_0$) and a GSM were developped for these campaigns and produced a lot of measurements at night time as well as during the day\cite{Aristidi05a, Aristidi09, Ziad08}. Taking advantage of the experience gained in developping these antarctic instruments, we propose now the Generalized Differential Image Motion Monitor, a compact instrument aiming at replacing the aging GSM. GDIMM is very similar to a DIMM, with 3 sub-apertures instead of 2. The seeing is obtained by the DIMM method, using two sub-pupils of same diameter. The third aperture has a diameter of 10~cm, allowing to estimate the isoplanatic angle via the scintillation. And the difference of the variances of the angle of arrivals through the pupils of different diameters makes it possible to derive the outer scale. The coherence time can be estimated from the temporal structure function of the angle of arrivals, thanks to the high sampling rate of the camera (up to 90~frames per seconds at full resolution). GDIMM has the advantage to be compact and easy to install, and is  controlled by a dedicated software. Efforts are in progress to make it fully automatic.

This paper is organized as follows. In section~\ref{par:theory} we briefly review the theory of the integrated turbulence parameters and the method used to estimate them. Section~\ref{par:instru}
presents the instrumental setup. Section~\ref{par:process} describes the
observations, the various calibration procedures, and the data processing. The results of the observations and a comparison with other instruments are presented
Sect.~\ref{par:results}. Conclusions and perspectives in Sect.~\ref{par:concl} end the paper.

%%%%%%%%%%%%%%%%%%%%%%%%%%%%%%%%%%%%%%%%%%%%%%%%%%%%%%%%%%%%%
\section{THEORY} 
\label{par:theory}

%%-----------------------------------------------------------
\subsection{Seeing} 
\label{par:seeing}
The seeing is the angular size of the FWHM of long-exposure images observed through the atmospheric turbulence. It is one of the most  important parameters describing the optical turbulence since it is related to the resolution of the images. Seeing monitors are operated in major observatories such as ESO Paranal, and  produce constant data used to optimize observations. The DIMM \cite{Sarazinroddier90, Verninmunoz95, Tokovinin02} is a seeing monitor which is very popular because of its simplicity. It is based on a small telescope with a entrance pupil made of 2 small subapertures, observing a bright single star with a short exposure (typically a few miliseconds). A tilt is given to the light propagating through one of the two apertures to produce twin images which move according to the turbulence. Angular distances $\Delta x$ and $\Delta y$ of the centroid of each star image are computed in both directions ($x$ is parallel to the basis of the sub-apertures). Longitudinal and transversal variances ($\sigma_l^2$ and $\sigma_t^2$ respectively) of $\Delta x$ and $\Delta y$ are estimated on a sequence of $N$ instantaneous snapshots ($N$ being  typically several hundreds). The seeing $\epsilon$ (in radian) and Fried parameter\cite{Fried66} $r_0$ (in $m$) are computed using the following formulae\cite{Tokovinin02}~:
\be
\epsilon_{l|t}=0.98\, (\cos z)^{-0.6}\frac{\lambda}{r_{0,l|t}}=0.98 \, (\cos z)^{-0.6}\:\left(\frac{D}{\lambda}\right)^{0.2}\:\left(\frac{\sigma_{l|t}^2}{K_{l|t}}\right)^{0.6}
\label{eq:seeing}
\ee
with
\begin{eqnarray}
K_l &=& 0.364\, (1-0.532 b^{-1/3}-0.024 b^{-7/3})\nonumber \\ \ \\
K_t &=& 0.364\, (1-0.798 b^{-1/3}+0.018 b^{-7/3})\nonumber
\end{eqnarray}
where $B$ is the distance between the sub-apertures, $D$ their diameter, $b=B/D$, $z$ is the zenithal distance and $\lambda$ the wavelength, traditionnaly set to 500~nm as a standard. Two estimations of the seeing are obtained for a given sequence, they are supposed to be the almost identical (isotropic hypothesis) and are averaged.

% 
%%-----------------------------------------------------------
\subsection{Isoplanatic angle} 
\label{par:isop}
The isoplanatic angle $\theta_0$ can be estimated from the scintillation of a single star observed through a pupil of diameter 10~cm and a central obstruction of~4~cm. The principle of the calculation is based on the similarity of the theoretical expressions of $\theta_0$ and the scintillation index $s$\cite{Looshogge79, Ziad00}. $\theta_0$ is obtained in arcsec for a wavelength $\lambda=500$~nm by the following formula
\be
\theta_0^{-5/3}=A \, (\cos z)^{-8/3}\, s
\label{eq:isop}
\ee
where $A=14.87$ is computed numerically from eqs. 19 and 21 of Ziad et al., 2000\cite{Ziad00}. $z$ is the zenithal distance of the star.

%%-----------------------------------------------------------
\subsection{Outer scale} 
The Von-Karman outer scale $\lo$ is related to the fluctuations of the angle of arrival of the light at a given position of the wavefront. The variance of the angular position of a star observed through a small aperture of diameter $D$ is given, in square radians by the following equation\cite{Ziad94}
\be
\sigma_D^2=0.17\, \lambda^2 \, r_0^{-5/3} \, (D^{-1/3}-1.525 \lo^{-1/3})
\label{eq:r0abs}
\ee
For isotropic turbulence, the variances in $x$ and $y$ directions are identical. The pupil of GDIMM has 3 apertures, two of diameter $D_1=6$~cm, one of diameter $D_3=10$~cm. The following ratio
\be
R=\frac{\sigma_{D_1}^2}{\sigma_{D_1}^2-\sigma_{D_3}^2}\; = \; \frac{D_1^{-1/3}-1.525 \lo^{-1/3}}{D_1^{-1/3}-D_3^{-1/3}}
\label{eq:l0}
\ee
makes it possible to estimate the outer scale $\lo$.

%%-----------------------------------------------------------
\subsection{Coherence time} 
The coherence time $\tau_0$ relevant for adaptive optics and interferometry, as defined by Roddier\cite{Roddier81} is
\be
\tau_0=0.31\, \frac{r_0}{\bar{v}}
\ee
where $\bar{v}$, the effective wind speed,  is a weighted average of the wind speed on the whole atmosphere\cite{Roddier81}. Ziad et al.\cite{Ziad12} have shown recently that it is possible to derive the effective wind speed from the the temporal structure functions $D_{x|y}(\tau)$ of the angle of arrivals, defined as
$$
D_{x|y}(\tau)=\langle \left(x|y(t)- x|y(t+\tau)\right)^2\rangle
$$
where $x$ (resp. $y$) stands for the angle of arrival (AA) in the $x$ (resp. $y$) direction (parrallel to the right ascension (resp. declination)). The brackets $\langle \rangle$ stand for temporal average. The AA is computed as the angular photocenter of the images produced by each sub-aperture. This function is zero for $\tau=0$ and saturates to a value $D_{\mbox{\scriptsize sat}}$ for $\tau\longrightarrow\infty$. We define its characteristic time $\tau_{a,x|y}$ as the value of $\tau$ for which
\be
D_{x|y}(\tau_{a,x|y})=\frac{D_{\mbox{\scriptsize sat}}}{e}
\ee
$\tau_{a,x|y}$ is indeed the AA coherence time. The effective wind speed as well as its direction $\gamma$ are derived from $\tau_{a,x}$ and $\tau_{a,y}$ using eqs.~10 and 11 of [\cite{Ziad12}], taking $k'=e$.

%%%%%%%%%%%%%%%%%%%%%%%%%%%%%%%%%%%%%%%%%%%%%%%%%%%%%%%%%%%%%
\section{INSTRUMENTATION} 
\label{par:instru}
%%-----------------------------------------------------------
\subsection{Telescope} 
\begin{figure}
\begin{center}
\includegraphics[width=10cm]{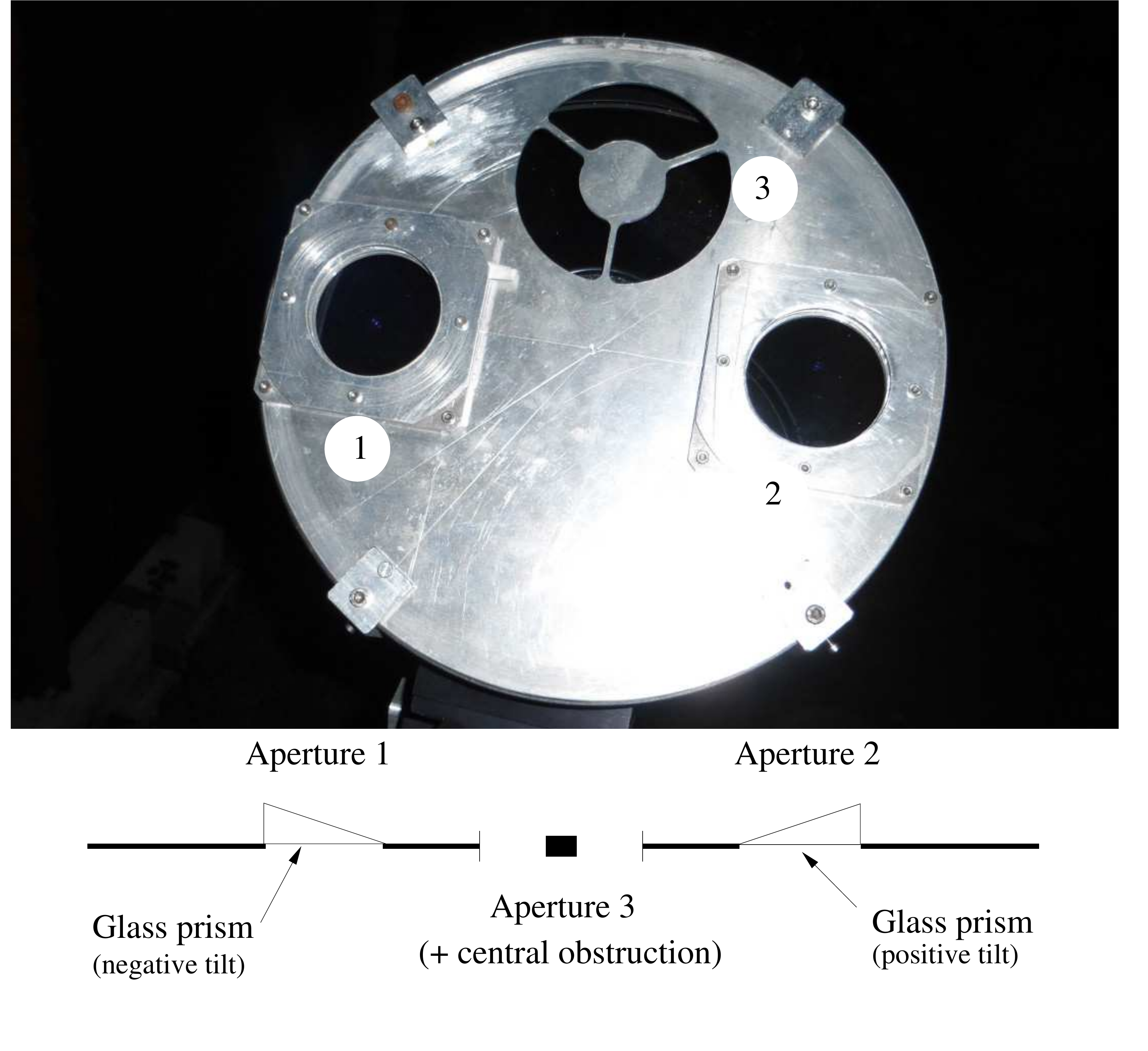}
\end{center}
\caption{Pupil mask of GDIMM. Top: photo taken in May 2014. Bottom: sectionnal view. Apertures 1 and 2 are circular with a diameter of 6~cm. They are both equipped with a glass prism to deviate the light away from the optical axis. Aperture 3 has a diameter of 10~cm and a central obstruction of 4~cm, it is used for isoplanatic angle and outer scale estimations. }
\label{fig:pupil}
\end{figure}

The telescope is a Schmidt-Cassegrain Celestron 11 (diameter 280~mm) equipped with an entrance mask with 3 sub-pupils as shown in Fig.~\ref{fig:pupil}. It is an extension of the classical 2-apertures DIMM mask, with a supplementary sub-pupil used for estimating the isoplanatic angle. Two sub-pupils are circular with a diameter of 6~cm, they are both equipped with a glass prism with a deviation angle of $\simeq 30$ arcsec. The prims are oriented to give opposite tilts to the incident light. The mak is oriented so that the aperture separation is parallel to the declination axis. The third sub-aperture is also circular, with  diameter 10~cm and a central obstruction of~4~cm to estimate the isoplanatic angle as described in section ~\ref{par:isop}. This aperture is left open and the corresponding image forms on the optical axis.
The telescope is placed on an Astro-Physics 900 equatorial mount controlled remotelly by a computer via a RS-232 link. The  tripod currently supporting the mount will be replaced in the future by a massive concrete pillar, allowing the pupil of GDIMM to be at a height of 5~m above the ground.

%%-----------------------------------------------------------
\subsection{Cameras} 
The main camera is a Prosilica EC650, with a CCD chip of 659$\times$493 pixels and a dynamic range of 12~bits. The spectral sensitivity spans across the whole visible range with a peak at a wavelength of 500~nm (the quantum efficiency at 500~nm is 50\%). The pixel size is $7.4 \mu$m$\times 7.4 \mu$m. The exposure time is adjustable by software from 10$\mu$s to 10~s, the maximum frame rate is 90~fps at full resolution (windowing and binning options are available to increase the frame rate if necessary). 

A Barlow lens increases the focal length to 7.8~meters to allow a slight oversampling of the Airy discs (5~pixels at a wavelength of 500~nm in the Airy disc of the 10~cm diameter sub-pupil). The camera is connected to the computer via a Firewire interface cable.

A second camera (USB webcam Logitech 9000) is placed at the focus of a finder with a wide field of 4 degrees. This camera is sensitive enough to detect bright stars.
%%-----------------------------------------------------------
\begin{figure}
\begin{center}
\includegraphics[width=10cm]{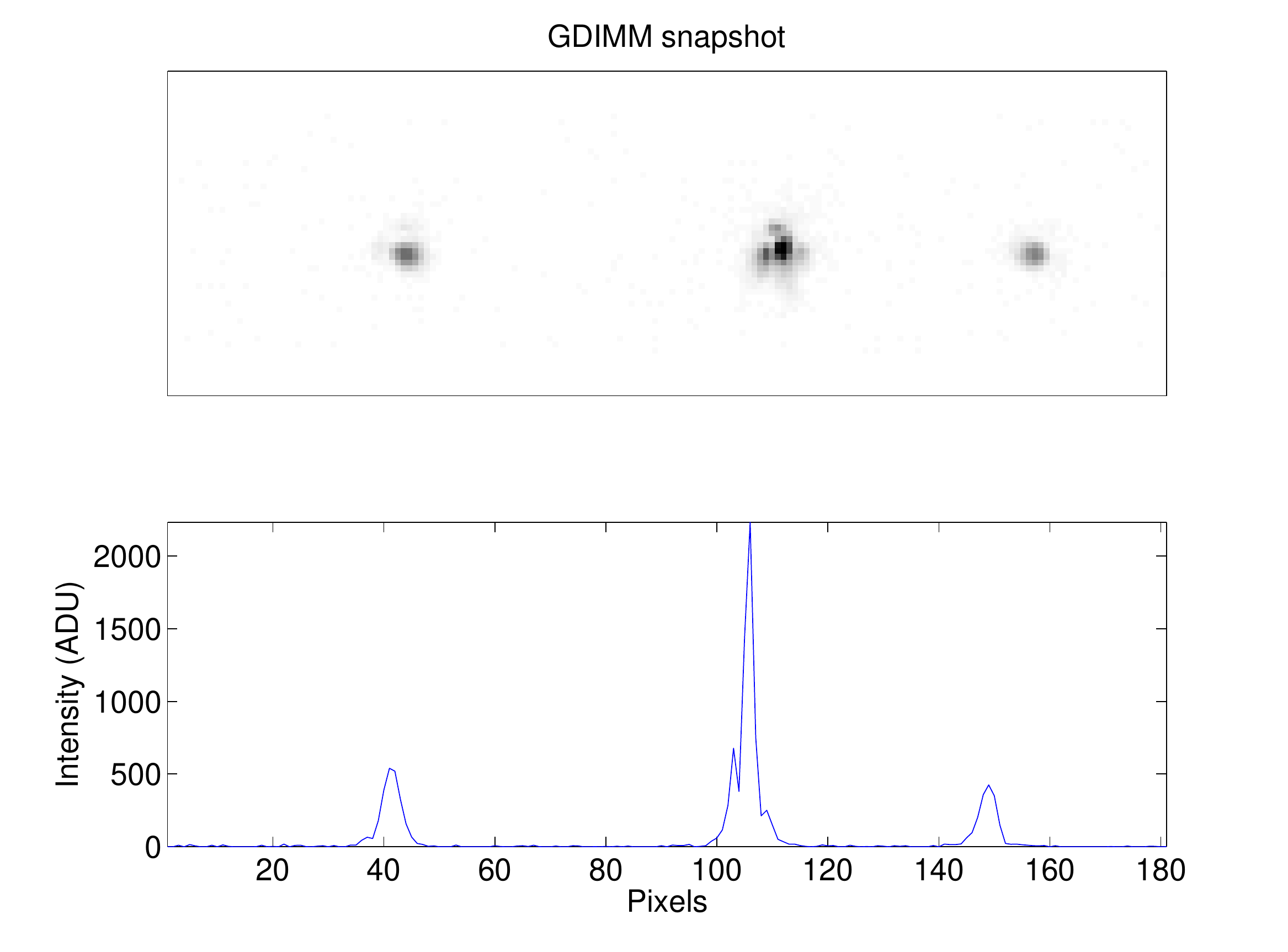}
\end{center}
\caption{GDIMM snapshot (zoom on the central part of the field). Top: grayscale plot showing the three images produced by the three subpupils. The central image corresponds to the 10~cm diameter aperture and is actually displaying speckles as the Fried parameter if lower than 10~cm. Bottom: horizontal cut of the image.}
\label{fig:snapshot}
\end{figure}

%%-----------------------------------------------------------
\subsection{Software} 
The acquisition software written in C++/QT takes benefit of years of developpement of automated instruments for the  Antarctic\cite{Aristidi05a, Daban10}. It can point the mount to the desired star, detects and centers automatically the target on the finder and on the science camera. Observations can be made manually or by acquisition sequences. Seeing and isoplanatic angle are computed in real time (in the future the coherence time and outer scale will be also given in real time). Various tests are performed to stop the observations when the star is lost (clouds) or if its zenithal distance becomes too large.

Data are written in text-based csv files as described in the next section. It is also possible to record the images in FITS cubes. The software has a ``simulation mode'': it can read a FITS cube from a previous run, and use it as input for a virtual observation. This is particulary useful for developping and reprocessing of the data after sofware improvement.

In the future we plan to install the GDIMM inside a dome at the Plateau de Calern and to remote-control the aperture of the dome via the acquisition software.

%%%%%%%%%%%%%%%%%%%%%%%%%%%%%%%%%%%%%%%%%%%%%%%%%%%%%%%%%%%%%
\section{OBSERVATIONS AND DATA PROCESSING} 
\label{par:process}
Observations with GDIMM are composed of continous sequences of one minute of time, each sequence giving one set of turbulence parameters: seeing $\epsilon$ (two values are calculated, longitudinal $\epsilon_t$ and transverse $\epsilon_l$), isoplanatic angle $\theta_0$, outer scale ${\cal L}_0$ and coherence time $\tau_0$. Currently the acquisition software computes in real time $\theta_0$ and the two values of $\epsilon$. Estimation of the outer scale is made afterwards since the algorithm is still in developpement, but we plan to include it to the software when it is stabilized. Coherence time calculation is not yet implemented.

A typical observing sequence is the following: 
\begin{itemize}
\item Every minute a cube of $N=1000$ continuous snapshots with a short exposure time $t$ of (3~ms gives sufficient signal to noise ratio for the brightest stars) is recorded into the computer memory (this takes about 20 seconds). 
\item Optionally a second cube of $N=1000$ images is recorded with a double exposure time $2t$ to compensate for finite exposure time. This option is denoted as ``2T mode''.
\item For each image of the sequence we compute the sky background on the upper edge of the field of view (about 1/10 of the field is considered), then substract it to the images and apply a threshold (5 times the standard deviation of the sky background).
\item For each frame we detect the three spots corresponding to the 3 sub-apertures, and compute their photocenters (in $x$ and $y$ directions) and their total flux.
\item A data file containing the UT acquisition time of each frame (with millisecond precision) spot photocenters and flux is generated for each cube.
\item Differential variances are computed from the photocenters of the two sub-images produced by the 6~cm diameter apertures, for the whole cube. In 2T mode, a compensation for exposure time is applied by the following formula\cite{Aristidi05a, Tokovinin02}
\be
\sigma^2_{l|t}(0)=\sigma^2_{l|t}(t)^{1.75}\, \sigma^2_{l|t}(2t)^{-0.75}
\ee
where $\sigma^2_{l|t}(t)$ (resp. $\sigma_{l|t}^2(2t)$) is the longitudinal$|$transverse differential variance computed for the exposure time $t$ (resp. 2t) and $\sigma^2_{l|t}(0)$ the longitudinal$|$transverse differential variance for zero~ms exposure time.
the coefficients 1.75 and 0.75 are taken from the paper of Tokovinin (2002)\cite{Tokovinin02}. Further comparison with the GSM is foreseen next summer to check these coefficients and adjust them if necessary to better fit our GDIMM data. 
\item The seeing (transversal and longitudinal)  is calculated from the variances using eq.~\ref{eq:seeing}. Scale calibration is made on bright double stars of known separation. Albireo (optical couple of separation 35$''$) or Mizar (very long period binary with separation of 14.5$''$) are good targets.
\item Scintillation index $s=\frac{\sigma_I^2}{\langle I\rangle^2}$ of the total flux $I$ of the central spot is computed for the $N=1000$ images of the cube. If the 2T mode is selected, the following formula is applied to compensate for the exposure time\cite{Ziad00}:
\be
s(0)=2 s(t)\, -\, s(2t)
\ee
\item The isoplanatic angle is calculated from eq.~\ref{eq:isop}
\item Variances, scintillation indexes, seeings and isoplanatic angle are summarized in a second data file, which contains, at the end of the observing run, all the integrated turbulence parameters for the night.
\end{itemize}

%%%%%%%%%%%%%%%%%%%%%%%%%%%%%%%%%%%%%%%%%%%%%%%%%%%%%%%%%%%%%
\section{RESULTS}
\label{par:results}
%%-----------------------------------------------------------
First test observations with GDIMM were carried out in June 2013 at the Plateau de Calern (South of France) on the bright stars Vega and Arcturus. The GSM instrument was operated simultaneously for cross-calibration. Results from this observing run are presented in this paper. More recently several test observations were performed on bright stars at the top of the Mont Gros (Nice, France). We present here data obtained on the nights of March 6th, March 20th and May 14th, 2014 on the star Arcturus.

\subsection{Seeing and isoplanatic angle}
\begin{figure}
\begin{center}
\includegraphics[width=7cm]{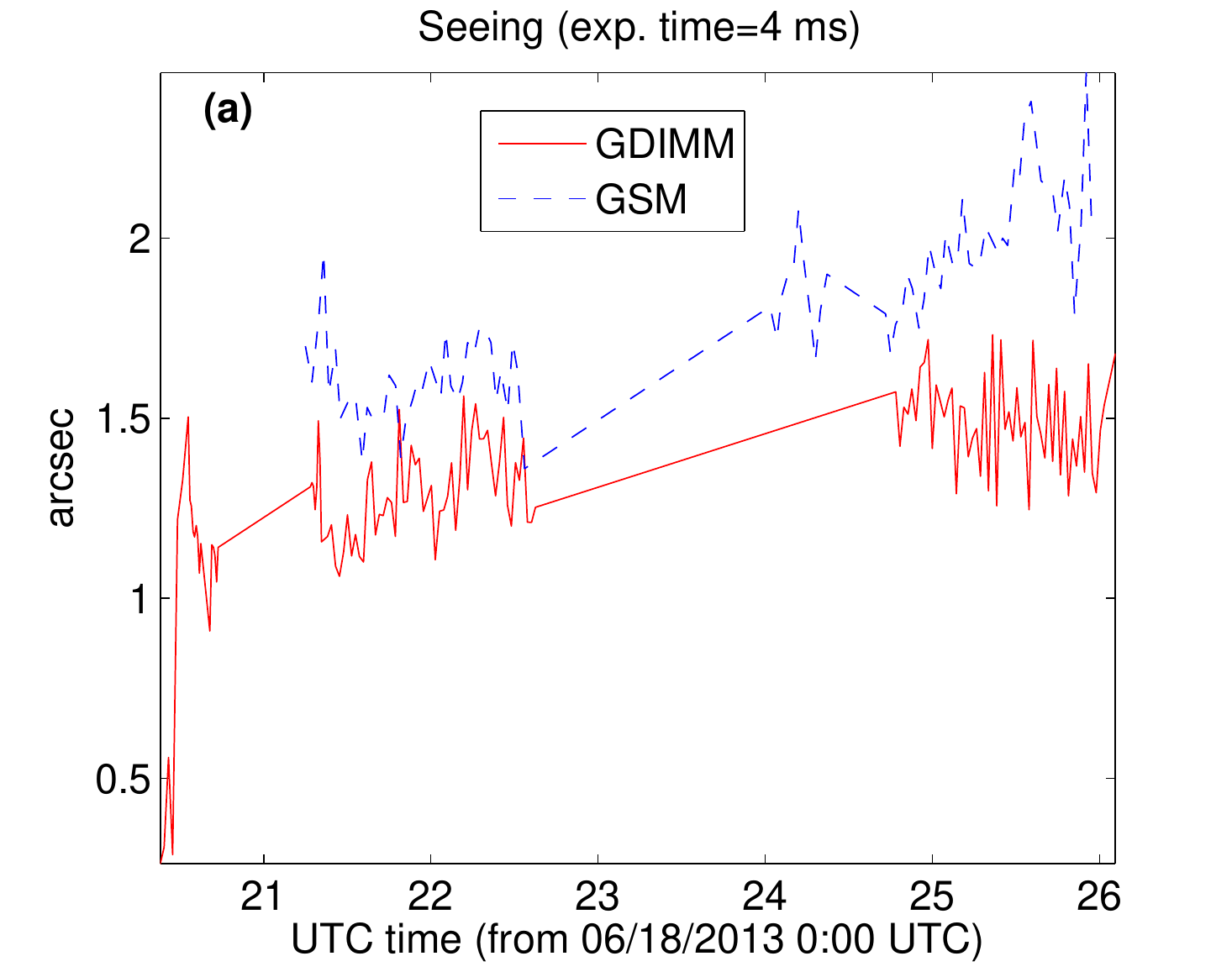} \ \ \includegraphics[width=7cm]{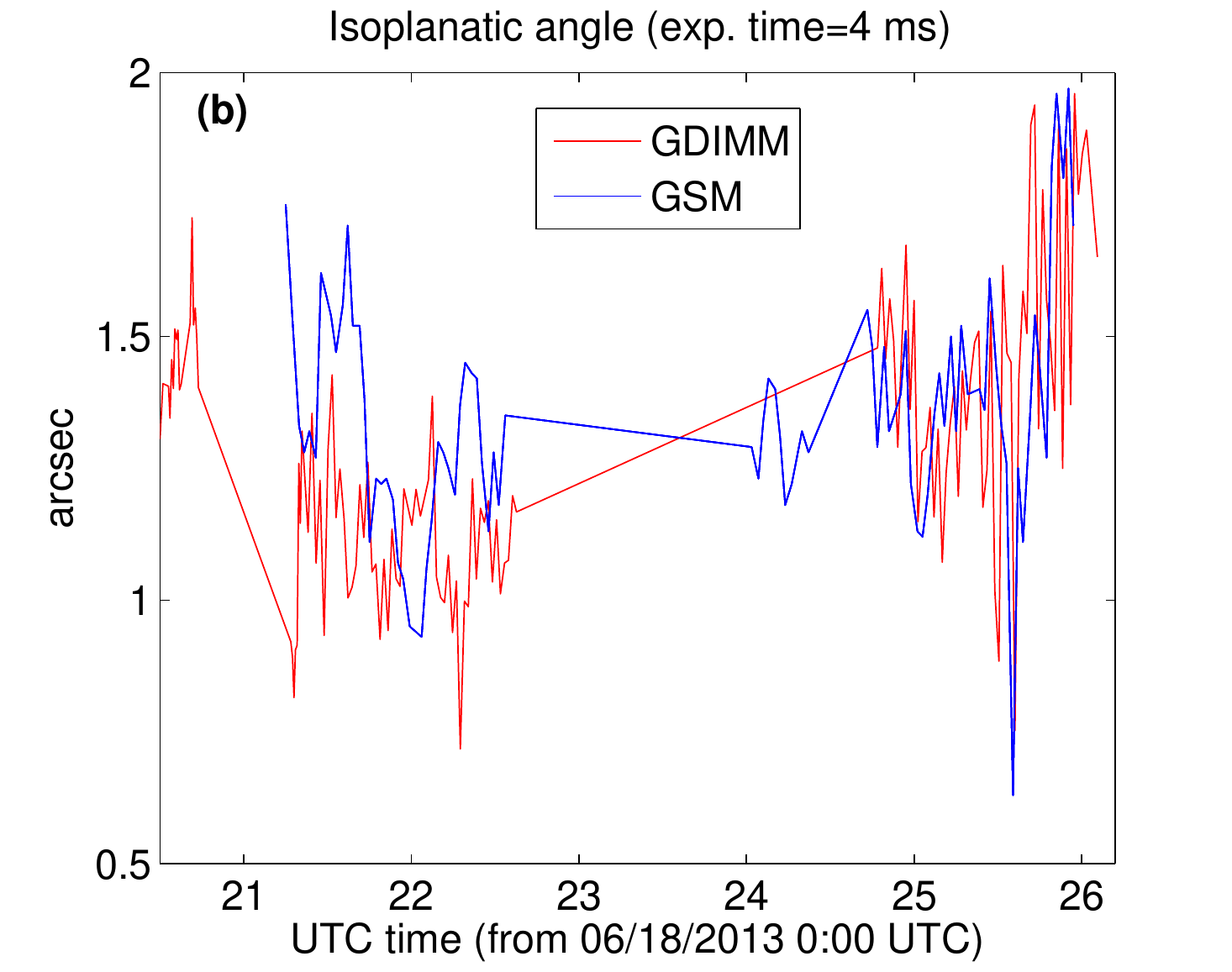} 
\end{center}
\caption{Times series of (a) the seeing and (b) the isoplanatic angle obtained on the night of June 18, 2013 at the plateau de Calern. Solid line: GDIMM data. Dashed line : GSM data (compensated from exposure time wheras GDIMM values are for an exposure time of 4~ms).}
\label{fig:seeing_isop_calern}
\end{figure}

Figure~\ref{fig:seeing_isop_calern}a presents times series of the seeing (average of transverse and longitudinal) obtained on June 18th 2013 at the Plateau de Calern. At that time the Barlow lens was not present and the compensation from exposure time was not implemented. Seeings are computed for an exposure time of 4~ms, and may be underestimated. Hence a small negative bias is observed between GSM and GDIMM seeings, but the shapes of the two curves are similar. The pixel size has not been calibrated with a double star for these data and we used the telescope focal to convert pixels into arcsec (this can explain part of the bias). The bias is increasing at the end of the sequence, probably due to a drop of the coherence time (meteo conditions were degradating). 

\begin{figure}
\begin{center}
\includegraphics[width=7cm]{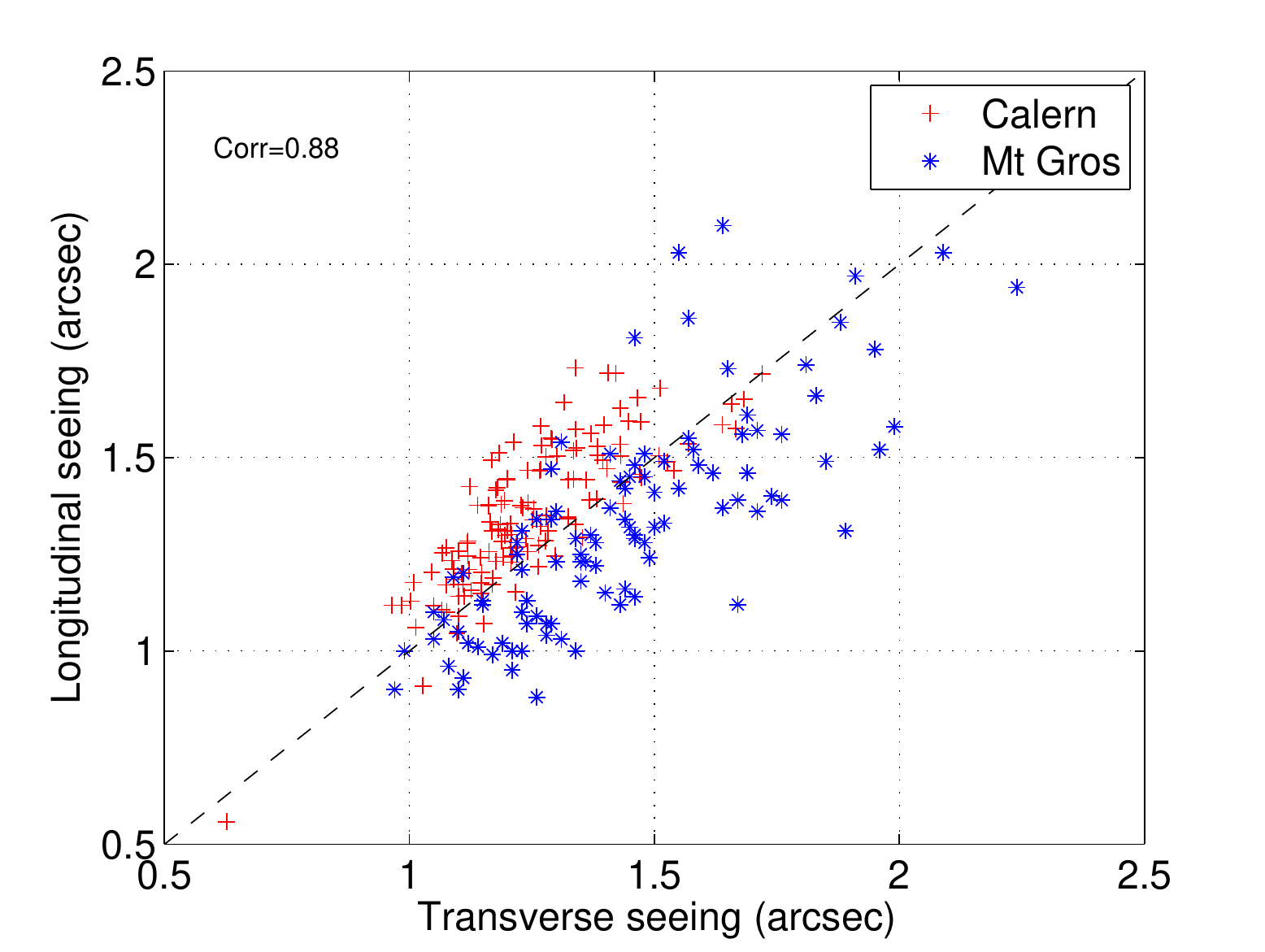} 
\end{center}
\caption{Transverse versus longitudinal seeings obtained with GDIMM obtained on the nights of June 18, 2013 at the Plateau de Calern and May 14, 2014 at the Mont Gros. In dashed-black the straight line of slope 1. The correlation coefficient of the combined data set is 0.88.}
\label{fig:seeing_lt}
\end{figure}

As mentionned in section~\ref{par:seeing}, the DIMM method is valid for isotropic turbulence, i.e. identical values of the transverse and longitudinal seeings $\epsilon_{t|l}$. Figure~\ref{fig:seeing_lt} shows a plot of $\epsilon_l$ versus $\epsilon_t$ for the two data sets. The cloud of points is spread along the ligne $\epsilon_t=\epsilon_l$ with small dispersion (a well known effects of the wind speed and direction and of the finite exposure time which was studied by Martin\cite{Martin87}). The correlation coefficient of 0.88 confirms the isotropic hypothesis. 

Isoplanatic angle times series for the run at the Plateau de Calern (June 18th 2013) obtained together with GSM is shown on fig.~\ref{fig:seeing_isop_calern}b. Very good agreement is found between the two data sets, despite the absence of compensation from finite exposure time.
%
%%-----------------------------------------------------------
\subsection{Outer scale} 
\begin{figure}
\begin{center}
\includegraphics[height=6cm]{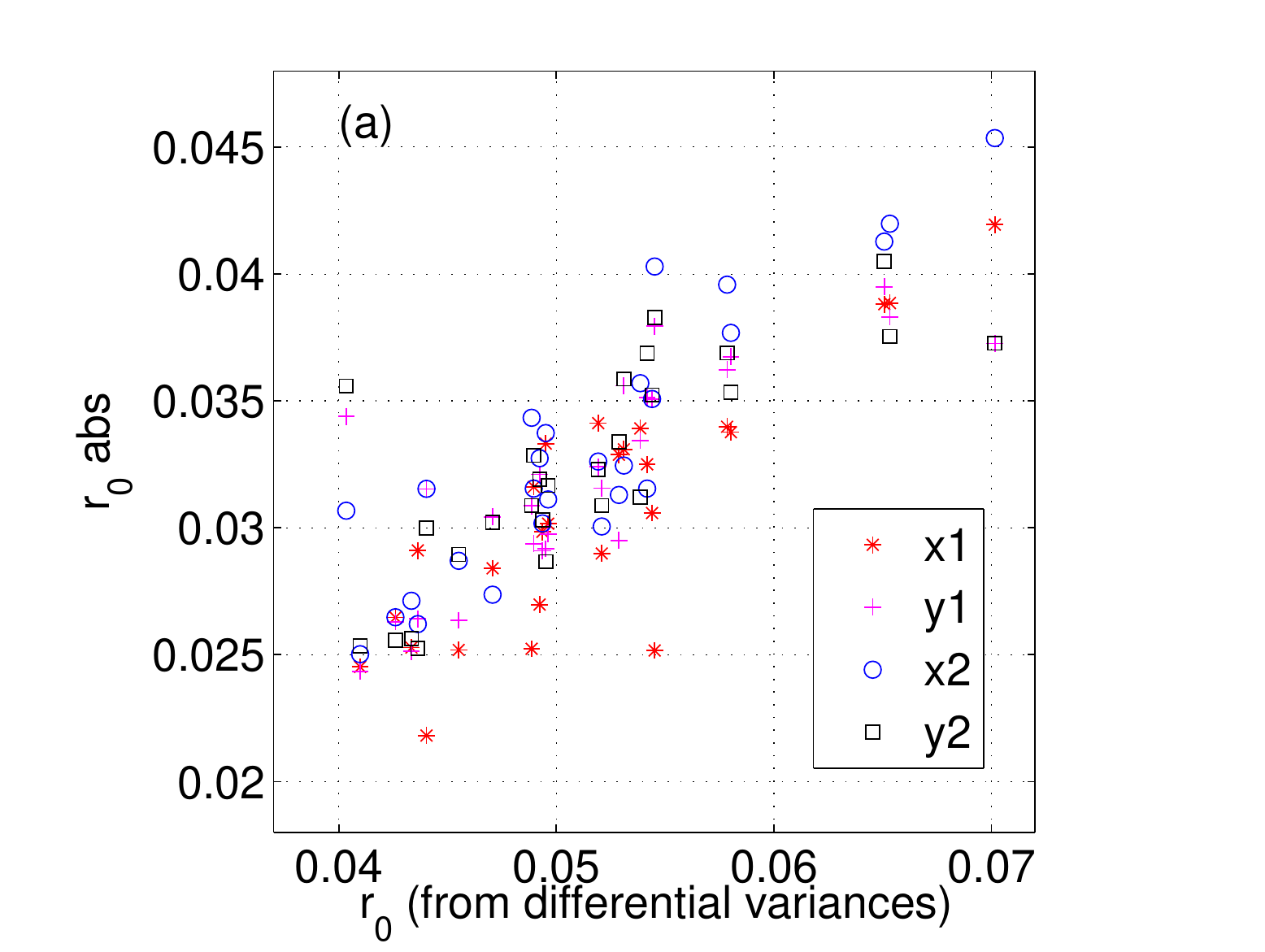} 
\includegraphics[height=6cm]{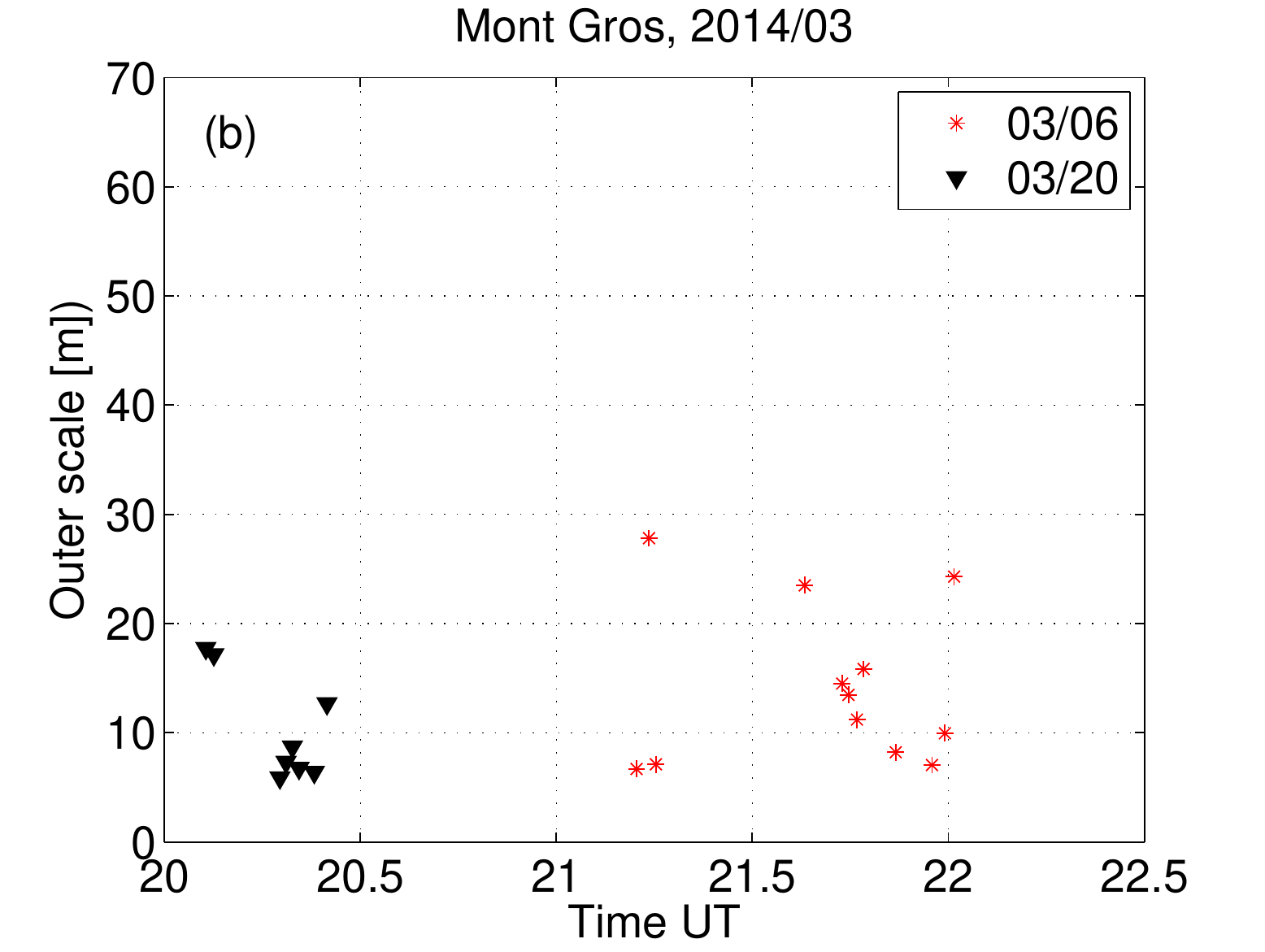} 

\end{center}
\caption{(a): Fried parameter $r_{0 \mbox{\scriptsize abs}}$ computed from the absolute variances $\sigma_{D_1}^2$, versus Fried parameter computed by the classical method using the differential variances. The 4 series of data correspond to the sub-pupils 1 and 2 in the $x$ and $y$ directions (namely $x_1$, $x_2$, $y_1$, $y_2$). (b): Time series of the outer scale $\lo$ (uncorrected from the exposure time) for data taken on March 6 and March 20, 2014 at the Mont Gros site.}
\label{fig:r0abs_r0diff}
\end{figure}

%%-----------------------------------------------------------
%
The outer scale is derived from the ratio $R$ defined in Eq.~\ref{eq:l0}. It requires the estimation of the ``absolute'' variances $\sigma_{D_1}^2$ and $\sigma_{D_3}^2$ of the angular position of the sub-images produced by the pupils of diameter $D_1=6$~cm  and $D_3=10$~cm. Typical values are $R\simeq 5$ for $\lo = 20$~m. Note that $R$ is independent of the seeing and the scale calibration. The drawback of this method is that $\sigma_{D_1}^2$ may contain some bias caused by telescope vibrations. Careful attention must be made during the processing in order to detect and eliminate contaminated data. We propose the following algorithm:
\begin{itemize}
\item From a data cube of $N$ frames we calculate 2 values of $\sigma_{D_3}^2$ from the photocenter of the instantaneous sub-images corresponding to the 10~cm diameter pupil, one in the $x$ direction and one in the $y$ direction.
\item Similarly we compute 4 values of $\sigma_{D_1}^2$ since we have two sub-pupils of diameter $D_1$.
\item From $\sigma_{D_1}^2$ it is possible to estimate a value $r_{0 \mbox{\scriptsize abs}}$ of the Fried parameter using eq.~\ref{eq:r0abs} and neglecting the term in $\lo$. Four values of $r_{0 \mbox{\scriptsize abs}}$ are computed for each cube and compared to the real value of $r_0$ calculated by the stantart way. We reject data for which $r_0-r_{0 \mbox{\scriptsize abs}}>\delta$ where $\delta$ is a threshold (we arbitrary took $\delta=4$~cm, but this is a point yet to be investigated). Fig.~\ref{fig:r0abs_r0diff}a displays a plot of the 4 values of $r_{0 \mbox{\scriptsize abs}}$ versus $r_0$ before applying the rejection,  for data obtained at the Mont Gros on March 6 and March 20, 2014. The high correlation coefficients (between 77\% and 90\%) indicate that the major part of the absolute variances was due to the turbulence for these data.
\item We finally compute 4 values of the ratio $R$ and derive 4 values of the outer scale for each cube. Outliers are rejeted, and the median of the remaining values is taken.
\end{itemize}

This algorithm was applied to data taken at the Mont Gros on the bright star Arcturus on March 6 and March 20, 2014. A total of 27 data cubes of $N=1000$ images with an exposure time of 5~ms was obtained during these two runs. After filtering (threshold and bad points), we kept 20 values of the outer scale $\lo$. Time series is presented on Fig.~\ref{fig:r0abs_r0diff}b. We found a mean outer scale $\lo\simeq 13$~m for the whole data set, with a good stability in time. These values are not compensated from the exposure time, they are indeed to be taken as a test of the method. 

%%-----------------------------------------------------------
\subsection{Coherence time} 
\begin{figure}
\begin{center}
\includegraphics[width=17cm]{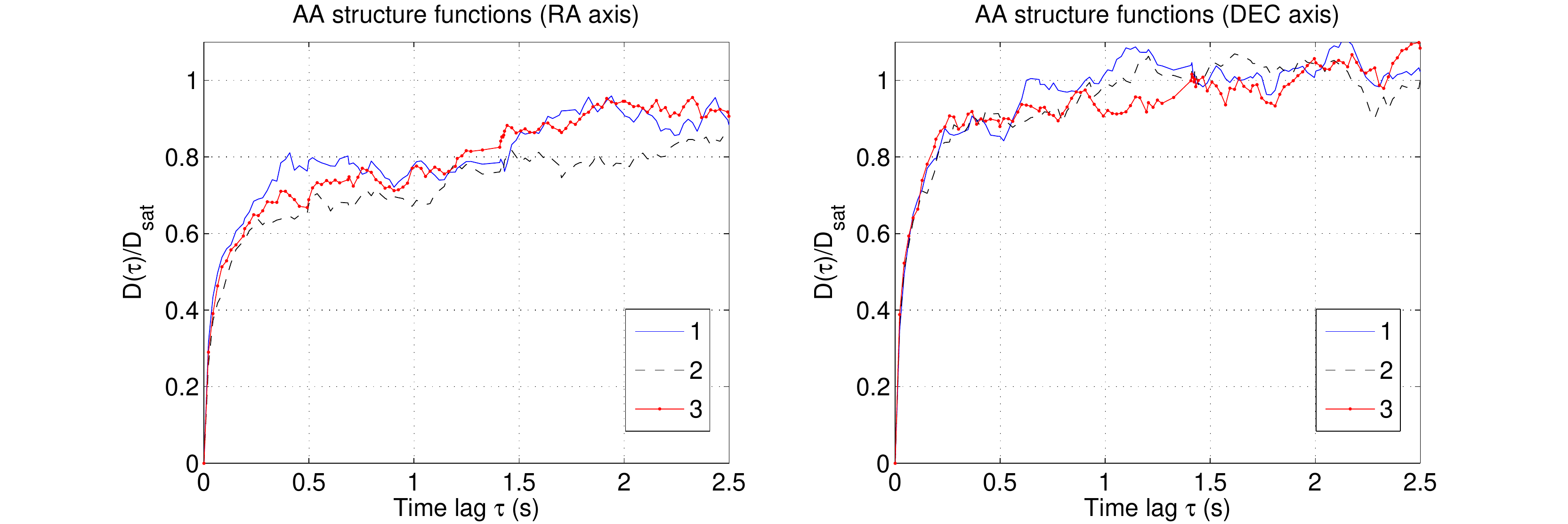}
\end{center}
\caption{Temporal structure functions $D_{x|y}(\tau)$ of the angle of arrivals for the data sequence recorded on May 14, 2014 at 20:09 UTC. $D_{x|y}$ is normalized to its saturation value. Left: curves for the AA in the $y$ direction (right ascension). Solid (1) and dashed (2) lines correspond to the subpupils of diameter 6cm, dotted line (3) corresponds to the 10~cm diameter aperture. Right: same curves for the $x$ direction (declination).}
\label{fig:temporal_st}
\end{figure}

%%-----------------------------------------------------------
\label{par:tau0}
The algorithm for the estimation of the coherence time is in developpement. It required the computation of the temporal structure functions $D_{x|y}(\tau)$ of the AA in both directions. Each sub-pupil gives an estimation of $D_{x}(\tau)$ and $D_{y}(\tau)$ so that we have 3 values of the AA coherence time $\tau_{a,x|y}$ in both directions. We show on fig.~\ref{fig:temporal_st} an example for a data cube obtained on May 14, 2014 at 20:01 UTC at the Mont Gros site. Structure functions for the $x$ (declination) direction saturate as expected when $\tau > 1$~sec. In the $y$ (right ascension) direction the function $D_{x}(\tau)$ almost saturates but there is a remaining trend probably due to telescope vibrations (pursuit of the star in the RA direction). 

In the $x$ direction we can estimate the AA coherence time to $\tau_{a,x} \simeq $22~ms. This value is not corrected from the exposure time (5~ms). This corresponds to an effective wind speed $\bar v \in [1.3 - 2.3]$~m/s depending on the wind direction (we took $\lo=20$~m but $\bar v$ is not very sensitive to $\lo$ and we found nearly the same result for $\lo=10$~m). The Fried parameter for this data cube, calculated by the DIMM method, is $r_0=5.4$~cm. This gives a coherence time $\tau_0\in [7 - 12]$~ms, which is of the order of magnitude of typical coherence times in the visible.

Currently a strong limitation of the precision of the method is the temporal sampling of the camera. It is 22~ms for these data, but we read the entire CCD for each acquisition. In the future we plan to use windowing options to reduce the time lag between successive snapshots.

%%%%%%%%%%%%%%%%%%%%%%%%%%%%%%%%%%%%%%%%%%%%%%%%%%%%%%%%%%%%%
\section{CONCLUSIONS AND PERSPECTIVES} 
\label{par:concl}
We have presented the GDIMM, a new turbulence monitor aiming at measuring the 4 integrated parameters of the optical turbulence ($\epsilon$, $\theta_0$, $\tau_0$ and $\lo$). The instrument is compact and aims at being a successor for the GSM. Seeing and isoplanatic angle monitoring are now completely functional. For the outer scale the algorithm is still in developpement. The first results are encouraging and give confidence in the possibility to stabilize a reliable monitoring in the next future. For the coherence time we just began to transpose the technique based on the structure functions presented by Ziad et al.\cite{Ziad12}. Section~\ref{par:tau0} shows that the method is applicable to our data, though precice estimation of $\tau_0$ will require to increase the frame rate of the acquisitions.

In the next months we plan to perform long-term observation campaigns at the plateau de Calern, together with GSM to finalize the GDIMM algorithms and test them in the widest possible range of situations. A GDIMM instrument is foreseen to be permanently installed on the top a 4~m high tower near the 1.5~m telescope MeO (plateau de Calern)\cite{Samain08}. It will give turbulence monitoring simultaneously with scientific MeO observations.

\section*{ACKNOWLEDGMENTS}      %>>>> equivalent to \section*{ACKNOWLEDGMENTS}       
 
We whish to thank Flavien Blary, Victorien Merl\'e and Paul Verdier who participated to the observations. Thanks also to Alex Robini for his valuable contribution to the instrument.

%%%%%%%%%%%%%%%%%%%%%%%%%%%%%%%%%%%%%%%%%%%%%%%%%%%%%%%%%%%%%
%%%%% References %%%%%

\bibliography{paper}   %>>>> bibliography data in report.bib

\begin{thebibliography}{10}

\bibitem{Ziad00}
Ziad, A., Conan, R., Tokovinin, A., Martin, F., and Borgnino, J. {\em Appl.
  Opt.}~{\bf 39},  5415 (2000).

\bibitem{Aristidi05a}
Aristidi, E., Agabi, A., Fossat, E., Azouit, M., Martin, F., Sadibekova, T.,
  Travouillon, T., Vernin, J., and Ziad, A. {\em A\&\/A}~{\bf 444},  651
  (2005).

\bibitem{Aristidi09}
Aristidi, E., Fossat, E., Agabi, A., M\'ekarnia, D., Jeanneaux, F., Bondoux,
  E., Challita, Z., Ziad, A., Vernin, J., and Trinquet, H. {\em A\&\/A}~{\bf
  499},  955 (2009).

\bibitem{Ziad08}
Ziad, A., Aristidi, E., Agabi, A., Borgnino, J., Martin, F., and Fossat, E.
  {\em A\&\/A}~{\bf 491},  917 (2008).

\bibitem{Sarazinroddier90}
Sarazin, M. and Roddier, F. {\em A\&\/A}~{\bf 227},  294 (1990).

\bibitem{Verninmunoz95}
Vernin, J. and Munoz-Tunon, C. {\em Pub. Astron. Soc. Pacific}~{\bf 107},  265
  (1995).

\bibitem{Tokovinin02}
Tokovinin, A. {\em Pub. Astron. Soc. Pacific}~{\bf 114},  1156 (2002).

\bibitem{Fried66}
Fried, D. {\em J. Opt. Soc. Am.}~{\bf 56},  1372 (1966).

\bibitem{Looshogge79}
Loos, G. and Hogge, C. {\em Appl. Opt.}~{\bf 18},  15 (1979).

\bibitem{Ziad94}
Ziad, A., Borgnino, J., Martin, F., and Agabi, A. {\em A\&\/A}~{\bf 282},  1021
  (1994).

\bibitem{Roddier81}
Roddier, F. {\em Progress in Optics}~{\bf 19},  281 (1981).

\bibitem{Ziad12}
Ziad, A., Borgnino, J., Dali-Ali, W., Berdja, A., Maire, J., and Martin, F.
  {\em J. Opt.}~{\bf 14},  045705 (2012).

\bibitem{Daban10}
Daban, J.~B. and al {\em SPIE Proc.} {\bf 7733},  77334T (2010).

\bibitem{Martin87}
Martin, H. {\em Pub. Astron. Soc. Pacific}~{\bf 99},  1360 (1987).

\bibitem{Samain08}
Samain, E. and al {\em Proc of the 16th International Workshop on Laser
  Ranging},  cddis.gsfc.nasa.gov/lw16, p.88 (2008).

\end{thebibliography}
\bibliographystyle{spiebib}   %>>>> makes bibtex use spiebib.bst

\end{document}